\documentclass[aps,10pt,showpacs,floatfix,citeautoscript,twocolumn,prb,superscriptaddress]{revtex4-1}

\usepackage{graphicx,xcolor}
\usepackage{amsmath}
\usepackage{comment}
\usepackage{textcomp}
\usepackage{inputenc}
\usepackage{ulem}

\newcommand*{\aas}{Ag$_{3}$AuSe$_{2}$}
\newcommand*{\aat}{Ag$_{3}$AuTe$_{2}$}
\newcommand{\agte}{Ag$_{2}$Te}
\newcommand{\aute}{Au$_{2}$Te}

\begin{document}

\title{High-pressure characterization of \aat: Implications for strain-induced band tuning}

\author{Juyeon Won}
\affiliation{Department of Materials Science and Engineering, University of Illinois at Urbana-Champaign, Urbana, IL 61801, USA}
\affiliation{Materials Research Laboratory, University of Illinois at Urbana-Champaign, Urbana, 61801 IL, USA}

\author{Rong Zhang}
\affiliation{Department of Applied Physics, Stanford University, Stanford, CA, 94305, USA}
\affiliation{Stanford Institute for Materials and Energy Sciences, SLAC National Accelerator Laboratory, Menlo Park, CA, 94025, USA}

\author{Cheng Peng}
\affiliation{Stanford Institute for Materials and Energy Sciences, SLAC National Accelerator Laboratory, Menlo Park, CA, 94025, USA}

\author{Ravhi Kumar}
\affiliation{Department of Physics, University of Illinois Chicago, Chicago, Illinois 60607, USA}

\author{Mebatsion S. Gebre}
\affiliation{Department of Materials Science and Engineering, University of Illinois at Urbana-Champaign, Urbana, IL 61801, USA}
\affiliation{Materials Research Laboratory, University of Illinois at Urbana-Champaign, Urbana, 61801 IL, USA}

\author{Dmitry Popov}
\affiliation{HPCAT, X-ray Science Division, Argonne National Laboratory, Lemont, Illinois 60439, USA}

\author{Russell J. Hemley}
\affiliation{Departments of Physics, Chemistry, and Earth and Environmental Sciences, University of Illinois Chicago, Chicago, Illinois 60607, USA}

\author{Barry Bradlyn} 
\affiliation{Department of Physics, University of Illinois at Urbana-Champaign, Urbana, 61801 IL, USA}
\affiliation{Materials Research Laboratory, University of Illinois at Urbana-Champaign, Urbana, 61801 IL, USA}

\author{Thomas P. Devereaux}
\affiliation{Department of Materials Science and Engineering, Stanford University, Stanford, CA, 94305, USA}
\affiliation{Stanford Institute for Materials and Energy Sciences, SLAC National Accelerator Laboratory, Menlo Park, CA, 94025, USA}

\author{Daniel P. Shoemaker}\email{dpshoema@illinois.edu}
\affiliation{Department of Materials Science and Engineering, University of Illinois at Urbana-Champaign, Urbana, IL 61801, USA}
\affiliation{Materials Research Laboratory, University of Illinois at Urbana-Champaign, Urbana, 61801 IL, USA}

%\date{\today}

\begin{abstract}
Recent band structure calculations have suggested the potential for band tuning in a chiral semiconductor, \aat, to zero upon application of negative strain. In this study, we report on the synthesis of polycrystalline \aat\ and investigate its transport, optical properties, and pressure compatibility. Transport measurements reveal the semiconducting behavior of \aat\ with high resistivity and an activation energy $E_a$ of 0.2 eV. The optical band gap determined by diffuse reflectance measurements is about three times wider than the experimental $E_a$. Despite the difference, both experimental gaps fall within the range of predicted band gaps by our first-principles DFT calculations employing the PBE and mBJ methods. Furthermore, our DFT simulations predict a progressive narrowing of the band gap under compressive strain, with a full closure expected at a strain of -4\% relative to the lattice parameter. To evaluate the feasibility of gap tunability at such substantial strain, the high-pressure behavior of \aat\ was investigated by \textit{ in situ} high-pressure X-ray diffraction up to 47 GPa. Mechanical compression beyond 4\% resulted in a pressure-induced structural transformation, indicating the possibilities of substantial gap modulation under extreme compression conditions.
\end{abstract}

\maketitle

\section{Introduction} 

Metal chalcogenides have emerged as a remarkable class of materials, exhibiting unique electronic properties that make them suitable for exploring topological phenomena.
Petzite or \aat\ is a mineral, discovered by Petz about two centuries ago\cite{faizan_carrier_2017}.
It belongs to the uytenbogaardtite mineral group along with \aas\ and {Ag$_{3}$AuS$_{2}$}. While \aas\ and \aat\ adopt the cubic $I4_132$ space group, {Ag$_{3}$AuS$_{2}$} takes on the tetragonal $P4_122$ structure.\cite{barton1978uytenbogaardtite}
Frueh et al.\cite{frueh_crystallography_1959} elucidated the crystal structure of natural \aat\ by single crystal X-ray diffraction (XRD). 

High-temperature powder XRD experiment revealed phase transitions in \aat\ at 210 and 319 $^{\circ}$C, with an intermediate phase eluding identification.\cite{cabri_phase_1965,frueh_crystallography_1959} These transitions were corroborated by differential thermal analysis.\cite{tavernier_uber_1967, smit_phase_1970}
%Below the melting point of \aat\ around 735 $^{\circ}$C, a continuous solid solution exists between the highest forms of \aat\ and \agte\ down to 319 $^{\circ}$C\cite{cabri_phase_1965}.
The initial synthesis of \aat\ was reported by Thompson,\cite{thompson_pyrosynthesis_1948} providing limited details on temperature or purity, except for conducting reactions in evacuated silica tubes. Young et al.\cite{young_thermoelectric_2000} investigated the thermoelectric properties of synthetic \aat\, noting low thermal conductivity and large electrical resistivity.

Recent computational predictions have pointed to a narrow and direct gap at $\Gamma$ \cite{faizan_elastic_2016,young_thermoelectric_2000,sanchez-martinez_spectral_2019}, opening possibilities for non-trivial topological behavior such as band crossing and inversion via band gap tuning.
S\'anchez-Mart\'inez et al.\cite{sanchez-martinez_spectral_2019} explored potential applications of \aas\ and \aat\ in dark matter detection, highlighting their tunable band gap through strain.
Subsequent band closure of \aat\ was predicted with a 4\% reduction in the lattice parameter,\cite{sanchez-martinez_spectral_2019} attainable through approaches such as chemical substitution and application of hydrostatic pressure.
Our preceding computational examination of the band gap tunability of \aas\ demonstrated a narrowing of the gap under tensile strain\cite{won_transport_2022}, aligning with findings by S\'anchez-Mart\'inez et al.\cite{sanchez-martinez_spectral_2019}
However, such heightened strain on a material is subject to deformation, decomposition, or fracture.
Therefore, a thorough assessment of the mechanical compressibility of \aat\ is necessary to evaluate the feasibility of band engineering by achieving -4\% strain.

Previous studies on \aat\cite{faizan_elastic_2016,young_thermoelectric_2000,sanchez-martinez_spectral_2019} highlight a small direct gap with potential band closure under compressive strain, yet experimental determination of the band gap and compressibility remains scant.
Here, we report the synthesis, transport, optical properties and mechanical compressibility of \aat\ and explore the DFT-simulated band structure at varying lattice constants.

\section{Methods}

%synthesis and characterization
Powders of Ag (99.9\% metals basis), Au (99.98\% metals basis), and Te (99.8\% metals basis) were loaded and sealed in an evacuated silica ampule in a 3.05:1:1.9 ratio. Subsequently, the ampule underwent annealing in a muffle furnace at 770 $^{\circ}$C for a day, followed by rapid cooling to 200 $^{\circ}$C and a cool-down to room temperature over three days. This procedure yielded a brittle ingot exhibiting a metallic luster and covered partially with gold flakes on the surface. High-resolution powder X-ray diffraction (PXRD) data were collected using the beamline 11-BM at the Advanced Photon Source (APS), Argonne National Laboratory (ANL).\cite{lee_twelve-analyzer_2008} Data were collected at 288 K, using a wavelength of 0.45897 \AA\ and the capillary geometry. Rietveld refinement was performed utilizing GSAS-II.\cite{toby_gsas-ii_2013}
%Additionally, cube-shaped crystals of \aat\ were picked out to confirm handedness and uniform chirality by single crystal X-ray diffraction (SCXRD). Transmission single crystal X-ray diffraction was measured in the \aat crystal using a Bruker D8 Venture diffractometer with Mo-K $alpha$ radiation at room temperature. Absorption correction was performed. The final refinements were done using SHELX.

% high-pressure XRD
High-pressure XRD measurements were performed at the 16-BMD-HPCAT beamline of the APS. Finely ground sample was loaded with a few grains of ruby and a Ne pressure-transmitting medium in a symmetric diamond anvil cell (DAC) with 300-$\mu$m culet diamonds and a Re gasket. The gaskets were indented to 60 $\mu$m, and a circular hole of 130 $\mu$m diameter was laser drilled to create the sample chamber. The Ne gas was loaded at the GSECARS facility at APS. The calibration of the sample center was carried out with a CeO$_2$ standard, and the high-pressure data were collected up to 47 GPa with an incident beam of wavelength 0.41326~\AA. Pressure at the sample region was measured and monitored by the shift of ruby fluorescence lines\cite{Mao_1986}. The XRD images were collected with a Pilatus detector and integrated using Dioptas software \cite{Clemens_2015}. 

%electrical resistivity
Electrical resistivity measurements were performed by the 4-point contact probe method using the Quantum Design Physical Property Measurement System (PPMS). The \aat\ ingot was cut and polished into a rod with dimensions 3.97 × 1.04 × 0.93 mm$^{3}$ to facilitate transport measurements. Gold leads were attached to the sample with a two-part silver epoxy for the measurements.
%gold lead diameter around 0.05 mm

%UV-Vis
Diffuse reflectance spectra of non-diluted \aat\ powder were acquired using a Varian Cary 5G UV-Vis-NIR spectrophotometer in the spectral range of 200-3000 nm. A Spectralon pellet, sourced from Harrick Scientific Products, served as the reference standard.
%A Tauc plot was generated from the collected reflectance data using Equation \ref{eq: bandgap}.
%Spectral bandwidth of 2 nm was used, and the light source was changed to 350 nm.

%DFT calculations
The density-functional theory simulations were performed employing the Vienna Ab initio Simulation Package (VASP) code\cite{vasp1993,vasp1994} with Perdew–Burke–Ernzerhof (PBE)\cite{PBE} projector-augmented wave (PAW) pseudopotentials.\cite{PAW1, PAW2}
We used the cutoff energy of 500 eV, and $\Gamma$ centered $8\times8\times8$ grid in the self-consistent field calculation. Calculations with modified Becke-Johnson (mBJ) exchange-correlation functions \cite{mBJ1,mBJ2} were performed with default parameters $\alpha = -0.012$, $\beta = 1.023$ bohr$^{1/2}$, $e = 0.500$ \cite{mBJ2}.
Experimental lattice constants in ambient pressure and compressive pressure were considered in the simulations.
The high-symmetry path in the band structure calculations is defined as $\Gamma (0.0, 0.0 0.0)$, $H (0.5, -0.5, 0.5)$, $N (0.0, 0.0, 0.5)$, $P (0.25, 0.25, 0.25)$.  

\section{Results and Discussion}

\subsection{Growth and structure}

\begin{figure}
\centering\includegraphics[width=\columnwidth]{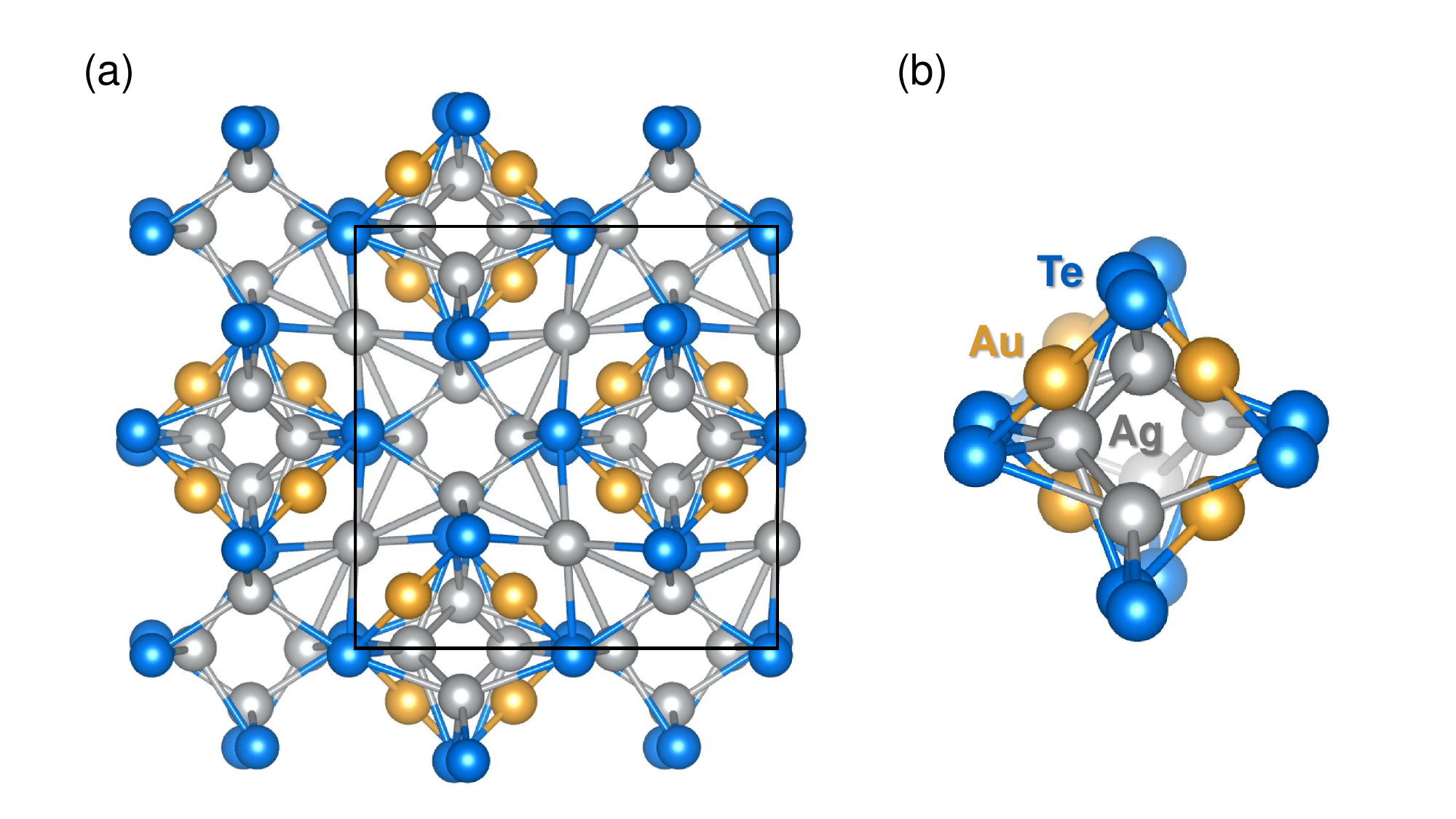} \\
\caption{The crystal structure of \aat\ is shown in (a) with the cubic unit cell indicated by the black box. The $4_1$ screw operation is well-demonstrated in (b) by the Ag chain spiraling into the plane.
}
\label{fig:aat-structure}
\end{figure}

\begin{figure}
\centering\includegraphics[width=\columnwidth]{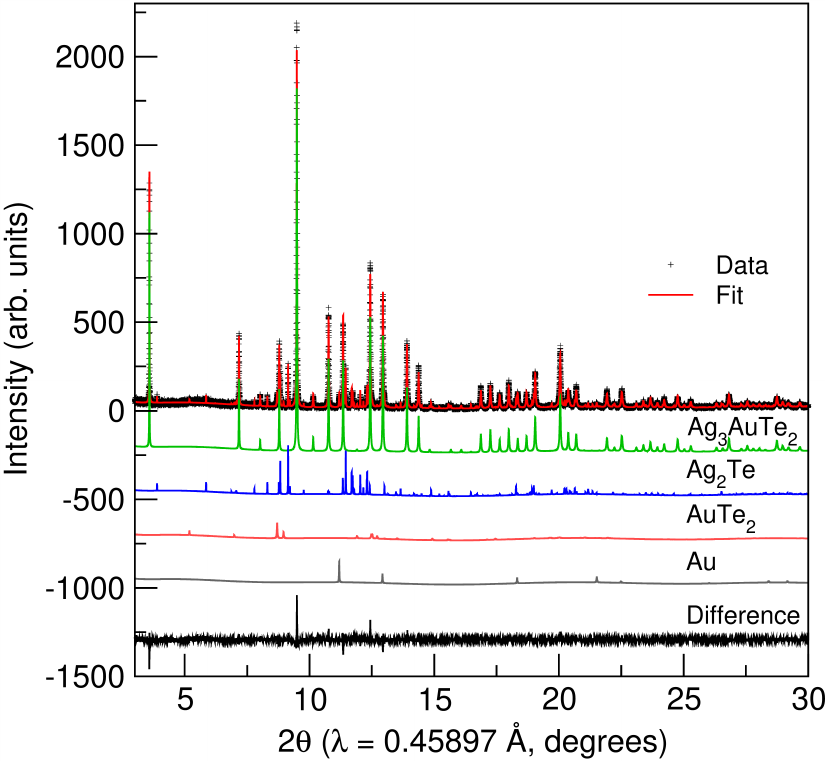} \\
\caption{
Rietveld-refined synchrotron PXRD pattern of \aat.
% obtained at the Advanced Photon Source beamline 11-BM.
The contributions from minor phases are shown in the figure and quantified in the text.
}
\label{fig:aat-pxrd}
\end{figure}

The room temperature crystal structure of petzite ($a = 10.38$ \AA, $Z = 8$) is analogous to fischesserite, \aas, as described in detail in our previous work.\cite{won_transport_2022}
%Figure \ref{fig:aat-structure} features the linear coordination of gold by two tellurium atoms and tetrahedrally coordinated silver by four tellurium atoms.
The crystal structure of \aat\ features a chiral $4_1$ Ag chain down the $\langle 100 \rangle$ axis, as illustrated in Figure \ref{fig:aat-structure}.
Our \aat\ sample was characterized by PXRD and analysed using Rietveld method.
As shown in Figure \ref{fig:aat-pxrd}, Bragg peaks were well-indexed by the cubic petzite structure and some minor impurities (wR = 2.84\%).
Phase fractions of \aat, \agte, Ag doped \aute\ and Au were refined to 0.926, 0.035, 0.034, and 0.005 by weight, respectively.

An equilibrated 3-component system can accommodate up to  three phases in an isotherm, which denotes an disequilibrium in our final product.
The Au-\agte\ phase diagram\cite{cabri_phase_1965} and previous studies on the thermal properties of \aat\cite{tavernier_uber_1967, smit_phase_1970,cabri_phase_1965,frueh_crystallography_1959} identify petzite as a line compound undergoing two phase transformations below its melting point.
%Of the two reported transitions, the intermediate phase around 200 C$^{\circ}$ remains unidentified and unindexed.
Despite our efforts to synthesize \aat\ via a slow cooling method from the melt, incorporating adjustments in cooling rates and elemental ratios, all attempts resulted in a mixture of \aat, \agte, \aute, and Au, contradicting the reported phase diagram.
There are certain ambiguities in the phase diagram proposed by Cabri et al. \cite{cabri_phase_1965}, such as the dashed line above the melting point of \aat\ labeled with a question mark.
In addition, it is unclear whether the "liquid" above the melting point is a uniform phase or molten impurities.
Furthermore, the \aat\ phase behavior near 200 $^{\circ}$C is ambiguous, with an intermediate phase mentioned but not indexed.
This prompts questions about potential phase separation or deviation from the nominal 312 stoichiometry, leading to phase separation around the transition temperature.
Consequently, the reformation of the lowest form of \aat\ upon cooling may be challenging due to the difficulty in reaching an equilibrium at such low temperatures.
\\

\begin{figure}[!htbp]
\centering\includegraphics[width=0.8\columnwidth]{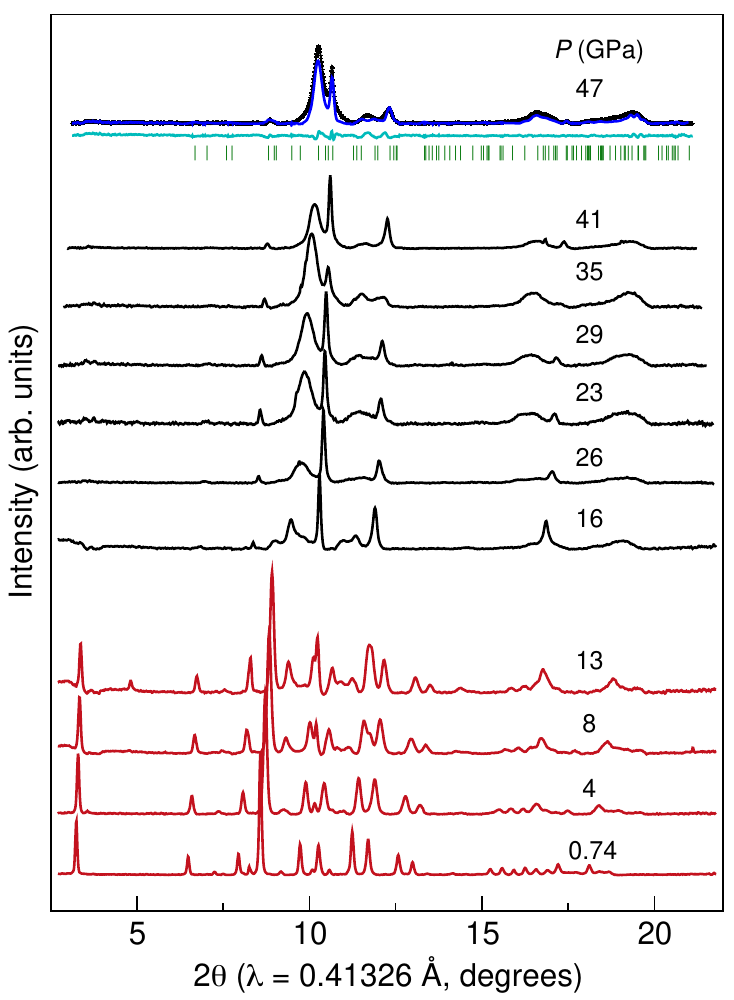} \\
\vspace{1em}
\centering\includegraphics[width=0.9\columnwidth]{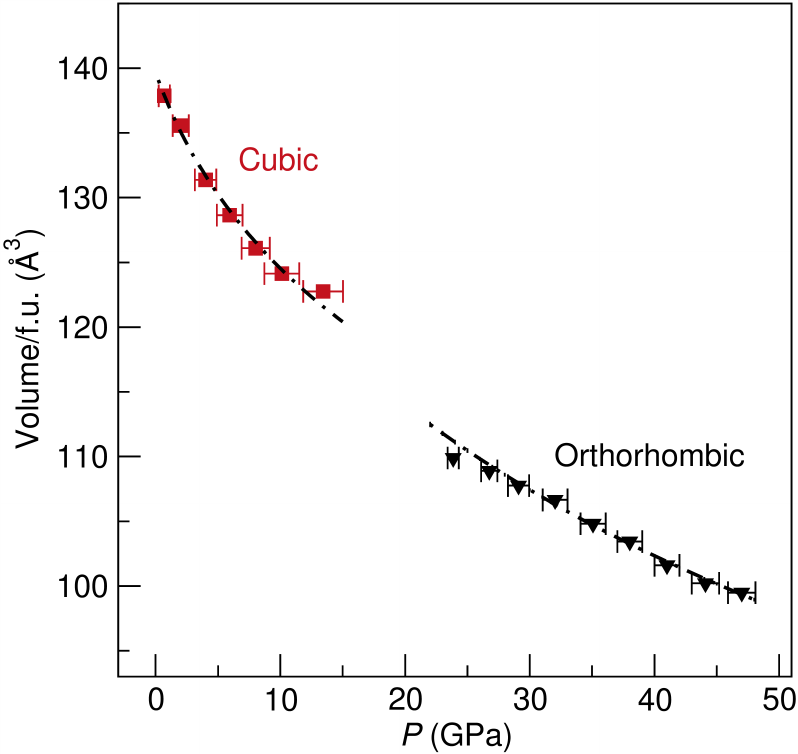} \\
\caption{
(Top) \textit{ in situ} high-pressure XRD patterns of \aat\ collected at various pressures ($\lambda = 0.41326$~\AA). Le Bail fit for the orthorhombic ($Pnma$) structure at 47 GPa is displayed. (Bottom) Volume per formula unit of refined cells for \aat\ plotted as a function of applied pressure. The multi-phase region around 16~GPa does not provide well-converged values for lattice parameters of the individual phases. The error bars for the volume are smaller than the symbol size.
}
\label{fig:aat-hpcat}
\end{figure}

\begin{figure}
\centering\includegraphics[width=\columnwidth]{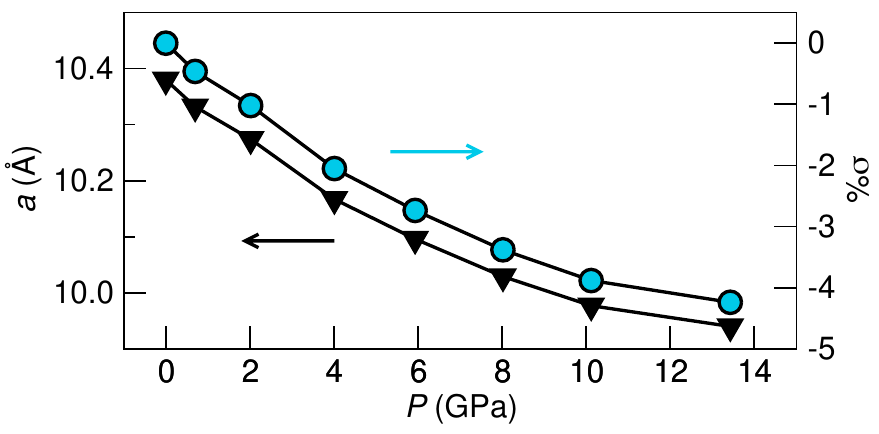} \\
\caption{Rietveld-refined lattice parameters and the corresponding $\sigma$ of compressed \aat\ are plotted against the applied pressure. Nearly -4\% of strain is achieved at 10 GPa.
}
\label{fig:a-vs-p}
\end{figure}

\subsection{High-Pressure X-ray Diffraction}

The high-pressure behavior of \aat\ was investigated through\textit{ in situ} high-pressure XRD experiments.
Figure \ref{fig:aat-hpcat} depicts the XRD patterns collected at various representative pressures, reaching up to 47 GPa.  
The cubic structure of \aat\ at ambient pressure remained stable up to 13 GPa, demonstrated by the rightward shift of the most intense (110) peak at 8.5$^{\circ}$.
Beyond 13 GPa, this prominent peak drops in intensity, indicating a pressure-induced phase change.
Notably, a number of peaks merge around 10$^{\circ}$ and new peaks emerge near 17.5$^{\circ}$.
Rietveld analysis confirmed the retention of the ambient cubic phase up to 13 GPa and the resulting refined lattice constant $a$ showed a gradual decrease with increasing pressure, reaching a maximum strain exceeding  -4\% at 13 GPa, as shown Figure \ref{fig:a-vs-p}. Even though the patterns collected between 13 and 20 GPa show a mixture of both low- and high-pressure phases of \aat, the significant changes observed in the diffraction peaks around 16 GPa imply structural transformation.
Analysis commenced using the 47 GPa pattern for the high-pressure phase corresponds mostly to the high-pressure phase.

We considered possible structural models from the pattern indexing and refinements using {\sc GSAS-II}, {\sc LHPM Rietica}, and {\sc QualX}\cite{toby_gsas-ii_2013, Hunter_2000, Altomare_2015} software programs as well as previously published papers for related compounds\cite{Saadi_2013}. 
Here, we consider an Al$_3$CePt$_2$-type hexagonal cell, $Pnma$ structure of CePd$_{3}$In$_{2}$,\cite{nesterenko_single_2004} and two high-pressure polymorphs of Ag$_2$Te: $C2/m$ from Zhu, et al.\cite{zhu_pressure_2016} and $Pnma$ from Zhang, et al.\cite{zhang_electronic_2015}
Attempts to fit the observed diffraction patterns with the above candidate structures favored the \agte-type structure, replicating most observed peaks at 47 GPa, as shown in Figure \ref{fig:aat-hpcat}.
The resulting cell parameters were a = 13.452(2) Å, b = 3.531(8) Å, c = 4.188(1) Å.
Fitting the patterns above 20 GPa with a similar cell resulted in a volume collapse of approximately 6\% around the phase transition, adding weight to the likeliness that \aat\ adopts the \textit{Pnma}-type symmetry at high pressure.
Ab initio enthalpy calculations on the proposed structural models would further shed light on their stability, especially since the number of observed diffraction peaks is small.

The bulk modulus of the low-pressure (cubic) and high-pressure (orthorhombic) phases was obtained by fitting the $P$-$V$ data to a third-order Vinet equation of state \cite{Vinet_1987}, as shown in Figure \ref{fig:aat-hpcat}.
The fits provided a bulk modulus value of $B_0$ = 52(7) GPa with a derivative $B_0'$ = 9.2 for the cubic phase and $B_0$ = 68(1) GPa with $B_0'$ = 4.8 for the orthorhombic phase.
The compressibility of \aat\ is comparable to other reported Ag ternary chalcogenides such as AgSbTe$_2$ and AgBiSe$_2$.\cite{Saadi_2013, Ko_2014, Elafy_2019}
The high strain tolerance of \aat\ bestow prospects for further investigation into the high-pressure band gap dynamics in light of the previous computational findings which suggested the possibility of strain-induced band engineering. 
Complementary techniques such as \textit{ in situ} high-pressure UV-Vis and resistivity measurements would provide valuable insights into its topological nature. 

\subsection{Electronic properties}

\begin{figure}
\centering\includegraphics[width=\columnwidth]{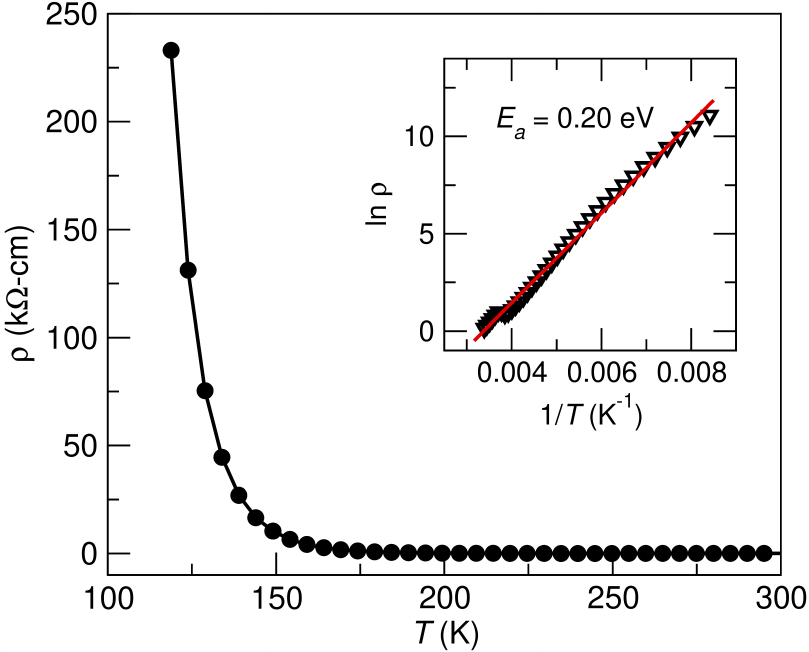} \\
\caption{
Electrical resistivity of \aat\ is plotted against temperature. The inset Arrhenius plot was obtained by applying a linear regression model with the least-squares method, which returned an activation energy of approximately 0.20 eV.
}
\label{fig: resistivity}
\end{figure}

Electronic properties of \aat\ were examined experimentally by transport measurements.
Figure \ref{fig: resistivity} shows the electrical resistivity of \aat\ as a function of temperature with the corresponding inset Arrhenius plot generated using the activation law.
The material exhibits a semiconductor behavior and high resistivity, nearing the megaohm range at around 100 K.
The magnitude of the activation energy $E_a$ aligns well with the previously proposed DFT-calculated band gaps.\cite{faizan_elastic_2016,sanchez-martinez_spectral_2019}
Young et al.\cite{young_thermoelectric_2000} reported similar experimental results, albeit with a slightly larger $E_a$ of 0.3 eV and no evidence for sample purity.
Differences in the estimated $E_a$ are within the typical error range for resistivity measurements stemming from factors such as imperfections in sample geometry, inhomogeneity, and possible in-gap defects.
Additionally, a resistivity kink near 275 K in Figure \ref{fig: resistivity}, also observed by Young et al.\cite{young_thermoelectric_2000} with no explanation, is likely an artifact from the freezing of adsorbed water.

Transport measurements confirm that \aat\ shows semiconductor-like temperature dependence and a band gap of at least 0.2 eV, accounting for potential defect hopping effects.
To evaluate petzite's potential for strain-induced band crossing, accurate determination as well as prediction of band gap is important.
Subsequent DFT analysis explores the band structure and possible band gap modulation through lattice parameter adjustments.

\begin{figure}
\centering\includegraphics[width=\columnwidth]{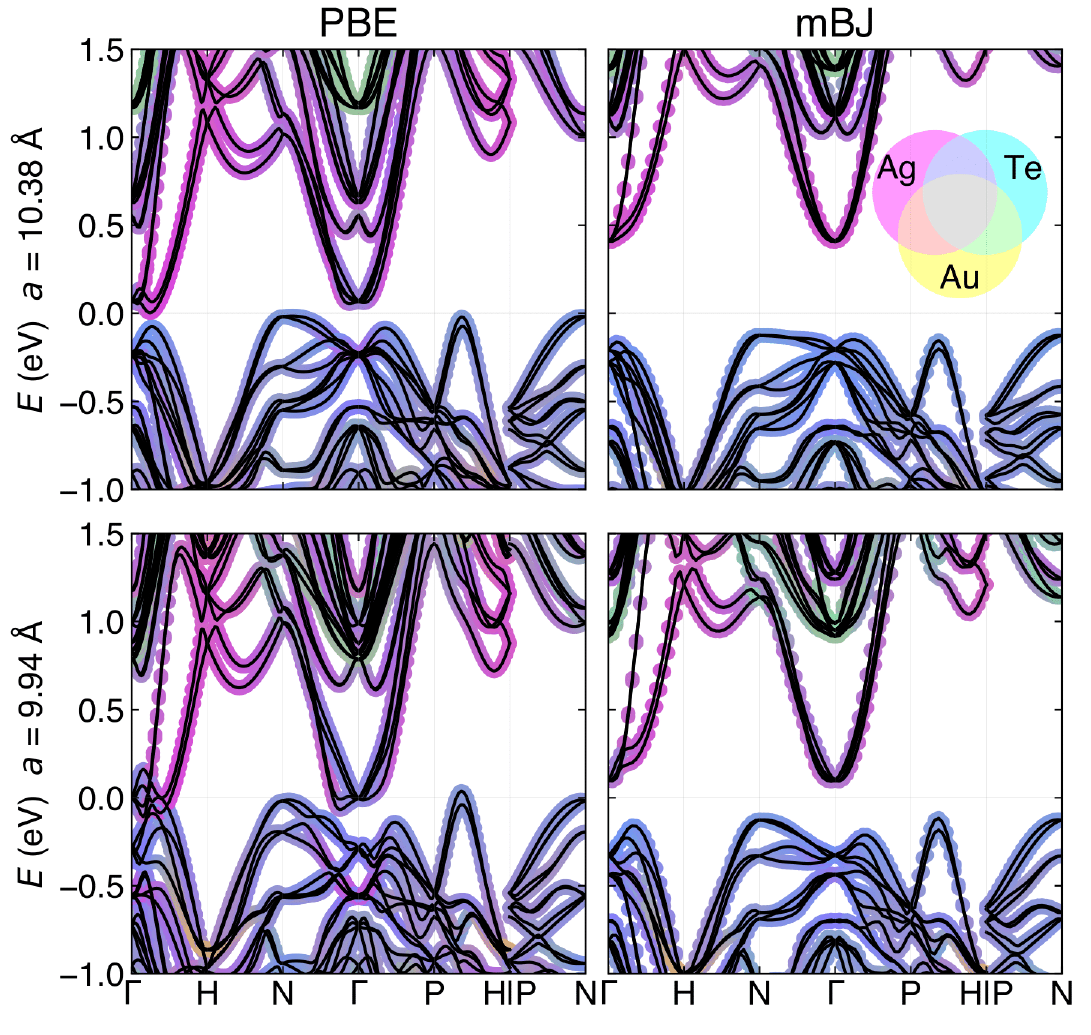} \\
\caption{
DFT-calculated band structures with spin-orbit coupling of \aat\ at ambient pressure and the maximum experimental compression before dissociation ($a=9.94$~\AA, 13~GPa). A semi-metallic behavior is predicted by the PBE functional, while the more accurate band gap predicting method, mBJ, predicts a gap even under pressure.
}
\label{fig:dft-bands}
\end{figure}

\begin{table}[]
\centering
\begin{tabular}{p{1cm}  p{1cm}  p{1cm}  p{1cm}  p{1cm}  p{1cm}}
\hline
\% &  0  &  1  & 2  & 3  & 4 \\
 \hline\hline 
PBE &  0.093   &  0.049   &  0.013  &  0  &  0  \\ 
mBJ &  0.603   &  0.516   &  0.438  &  0.383  &  0.272 \\
\hline
\end{tabular}
\caption{\label{tab: DFT-gaps} DFT-calculated band gaps of \aat\ with an incremental reduction in the lattice parameter. Band gaps are in the unit of $\text{eV}$. The PBE and mBJ gaps decrease with reduced lattice parameters, but at 4\% compressive strain, the PBE gap achieves metallicity while the mBJ gap remains open.
}
\end{table}

Electronic properties were computationally investigated by density functional theory (DFT) band structure calculations.
Figure \ref{fig:dft-bands}\ illustrates the calculated band structure of \aat\ with a reduction of lattice constant up to 4\% compressive strain, using the Perdew-Burke-Ernzerhof parameterization (PBE) and modified Becke-Johnson (mBJ), including spin-orbit coupling (SOC).
Calculations without SOC are provided in the supplementary material.
SOC breaks Kramer's degeneracy and splits the band structure for different spin components, but it does not induce qualitative differences in the band dispersion and the gap.
Table \ref{tab: DFT-gaps} summarizes the results, indicating that at ambient pressure ($a = 10.38$~\AA), the indirect band gap between the valence band near $N$ and the conduction band around $\Gamma$ is 0.093 and 0.603 eV for PBE and mBJ functionals, respectively.
mBJ calculations predict that negative strain (compression) decreases the band gap while conserving the overall shape and features of the band structure, and the material remains to be an insulator at the maximal compressive strain (-4\%).
The PBE functional, known to underestimate band gaps, also predicts decreasing band gaps with negative strain.
However, the conduction and valence band overlap near $\Gamma$ along the $\Gamma - H$ cut before reaching -4\% strain, transitioning the material into a metal under pressure.
The SOC splitting strengthens with increased strain and the predicted SOC effect by mBJ functional is less significant than that using PBE method.
Our PBE results are mostly consistent with previous DFT studies\cite{sanchez-martinez_spectral_2019, Martinez2023}.
However, we expect the mBJ functional to provide more accurate calculation of the band structure in general, and the shape and splitting of the conduction band around $\Gamma$, in particular, are qualitatively different from the PBE predictions.

To unambiguously visualize the atomic contribution near the band edges,  Ag, Te, and Au are color-coded by magenta, cyan, and yellow, respectively, in Figure \ref{fig:dft-bands} for PBE and mBJ functionals on strained and ambient lattices.
The band structures are highly hybridized. The valence bands are primarily composed of an equal mixture of Ag-d and Te-p orbitals, indicated by blue shades, and a smaller contribution from Ag-s orbitals.
The conduction bands largely consist of Au-s and d orbitals, although contributions from other atoms remain substantial.
In general, the band character of the conduction bands becomes mixed with the valence bands as the band gap is reduced. 

Ag$_3$AuTe$_2$, along with the isostructural Ag$_3$AuSe$_2$, were theoretically analyzed in Refs.~\onlinecite{sanchez-martinez_spectral_2019} and \onlinecite{won_transport_2022}. In particular, Ref.~\onlinecite{sanchez-martinez_spectral_2019} used DFT calculations to predict that both compounds feature a gap that narrows as a function of applied pressure. The low-energy bands under pressure are centered at the $\Gamma$ point and involve fourfold degenerate chiral multifold fermions with nontrivial Chern numbers. Ref.~\onlinecite{sanchez-martinez_spectral_2019} developed a low-energy $\mathbf{k}\cdot\mathbf{p}$ approximation of the bands near $\Gamma$ and showed that in Ag$_3$AuSe$_2$ the optical conductivity under pressure was dominated by contributions from the multifold fermions~\cite{bradlyn2016beyond,flicker2018chiral}. Due to the qualitative similarity between the electronic bands of Ag$_3$AuSe$_2$ and Ag$_3$AuTe$_2$, we expect that optical properties of Ag$_3$AuTe$_2$ are also due to the multifold fermions.

\subsection{Optical Properties}

\begin{figure}
\centering\includegraphics[width=\columnwidth]{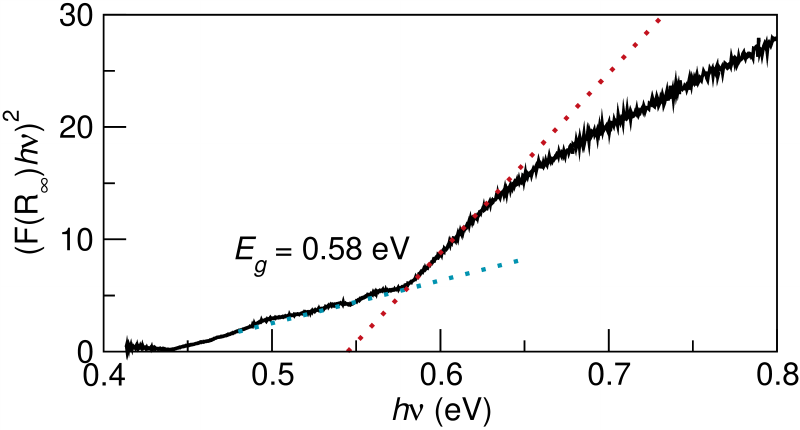} \\
\caption{
Tauc plot of \aat\ was obtained from the diffuse reflectance measurement. The optical band gap energy was estimated by the intersection of the baseline (blue) and fitted line (red) since the baseline was not flat to zero. The approximate band gap is 0.58 eV.
}
\label{fig:aat-tauc}
\end{figure}

Optical characterization of \aat\ was performed by UV-Vis diffuse reflectance measurements at room temperature.
The obtained reflectance spectra were then transformed into absorption spectra using the Kubelka-Munk function,\cite{kubelka_beitrag_1931} expressed as
 
%   (\alpha\cdot\\h\nu)^{1/\gamma} = B(h\nu-E_g)
%   \label{eq:tauc},
%\end{equation}

\begin{equation}
   F(R_\infty) = \frac{(1-R_\infty)^2}{2R_\infty}
   \label{eq:kubleka-munk},
\end{equation}
where $R_\infty$ represents the reflectance of a medium with infinite thickness.
Replacing the absorption coefficient $\alpha$ in the Tauc equation\cite{tauc_optical_1966} with $R_\infty$ yields the expression (2) 

\begin{equation}
   (F(R_\infty)\cdot\\h\nu)^{1/\gamma} = B(h\nu-E_g)
   \label{eq:bandgap},
\end{equation}
where the $\gamma$ factor is equal to 1/2 for a direct gap and 2 for an indirect gap.

Using the above relations with $\gamma = 1/2$ resulted in a Tauc plot with a gentle, rather than a steep rise at approximately 0.6 eV, as shown in Figure \ref{fig:aat-tauc}, consistent with the predicted mBJ ambient pressure indirect gap.
However, the PBE gap falls short of the optical band gap by orders of magnitude, as expected due to the tendency of PBE functional to underestimate band gaps.
The Arrhenius activation energy $E_a$ (0.2 eV) is lower than but comparable to the optical gap in magnitude.
Discrepancies in the experimental gap values may arise from factors such as errors in sample dimensions for transport measurement and a rough approximation of the optical absorption onset. 
The correspondence between the optical and mbJ gap suggests promising avenues for exploiting pressure-induced band modulation of \aat.

\section{Conclusions}

We investigated the synthesis, electronic and optical properties, as well as the compressibility of polycrystalline \aat.
High-pressure XRD measurements revealed a pressure-induced structural transition of petzite around 16 GPa, shifting from a cubic to an orthorhombic Ag$_2$Te-type structure.
Transport results corroborated Young et al.'s observations,\cite{young_thermoelectric_2000} highlighting the semiconductor nature of the telluride with an Arrhenius activation energy of 0.2 eV, which falls within the predicted gap range by PBE and mBJ DFT calculations.
The absorption band edge determined by diffuse reflectance measurements was in excellent agreement with the mBJ gap.
First-principles calculations using PBE and mBJ functionals predicted a gap reduction with decreasing lattice parameters, reaching zero and 0.3 eV gap at -4\% strain, respectively.
Our high-pressure XRD experiments demonstrated a lattice parameter reduction of over 4\% in \aat, suggesting a viable band engineering avenue for \aat\ towards zero gap under applied pressure.

\section{Acknowledgments}

Crystal growth, transport, and optical characterization were supported by the Center for Quantum Sensing and Quantum Materials, an Energy Frontier Research Center funded by the U.\ S.\ Department of Energy (DOE), Office of Science (SC), Basic Energy Sciences (BES) under Award DE-SC0021238. First-principles calculations conducted by RZ and CP were supported by the DOE-SC-BES under Award DE-AC02-76SF00515. The authors acknowledge using facilities at the Materials Research Laboratory Central Research Facilities, University of Illinois, partially supported by NSF through the University of Illinois Materials Research Science and Engineering Center (DMR-1720633). High-pressure X-ray diffraction measurements were performed at HPCAT (Sector 16), Advanced Photon Source (APS), Argonne National Laboratory. HPCAT operations are supported by the DOE-National Nuclear Security Administration (NNSA) Office of Experimental Sciences. The beamtime and contributions of RK and RJH were made possible by the Chicago/DOE Alliance Center (CDAC), which is supported by DOE-NNSA (Grant No. DE-NA0003975). RK and RH were also supported by NSF 
DMR-2119308 and DMR-2104881. COMPRES supported use of the gas loading system under NSF Cooperative Agreement No. EAR-1606856 and by GSECARS through NSF Grant No. EAR-1634415 and DOE Grant No. DE-FG02-94ER14466. The Advanced Photon Source is a DOE Office of Science User Facility operated for the DOE Office of Science by Argonne National Laboratory under Contract No. DE-AC02-06CH11357. This research used the National Energy Research Scientific Computing Center, a DOE Office of Science User Facility supported by the Office of Science of the U.S. Department of Energy under Contract No. DE-AC02-05CH11231 using NERSC award BES-ERCAP0027203.

\bibliographystyle{apsrev4-1}
\bibliography{aat}

\end{document}

%\maketitle

\begin{center}
\Large 
\textbf{High-pressure characterization of \aat
: Implications for strain-induced band tuning}\\
\vspace{1em}
Supplementary Material\\
\vspace{1em}
\normalsize
Juyeon Won, Rong Zhang, Cheng Peng, Ravhi Kumar, Mebatsion S. Gebre, Dmitry Popov, Russell J. Hemley, Barry Bradlyn, Thomas P. Devereaux, Daniel P. Shoemaker

\vspace{2em}
\end{center}

\subsection*{Band Structure Properties}

\begin{figure}[h]
\centering\includegraphics[width=1\columnwidth]{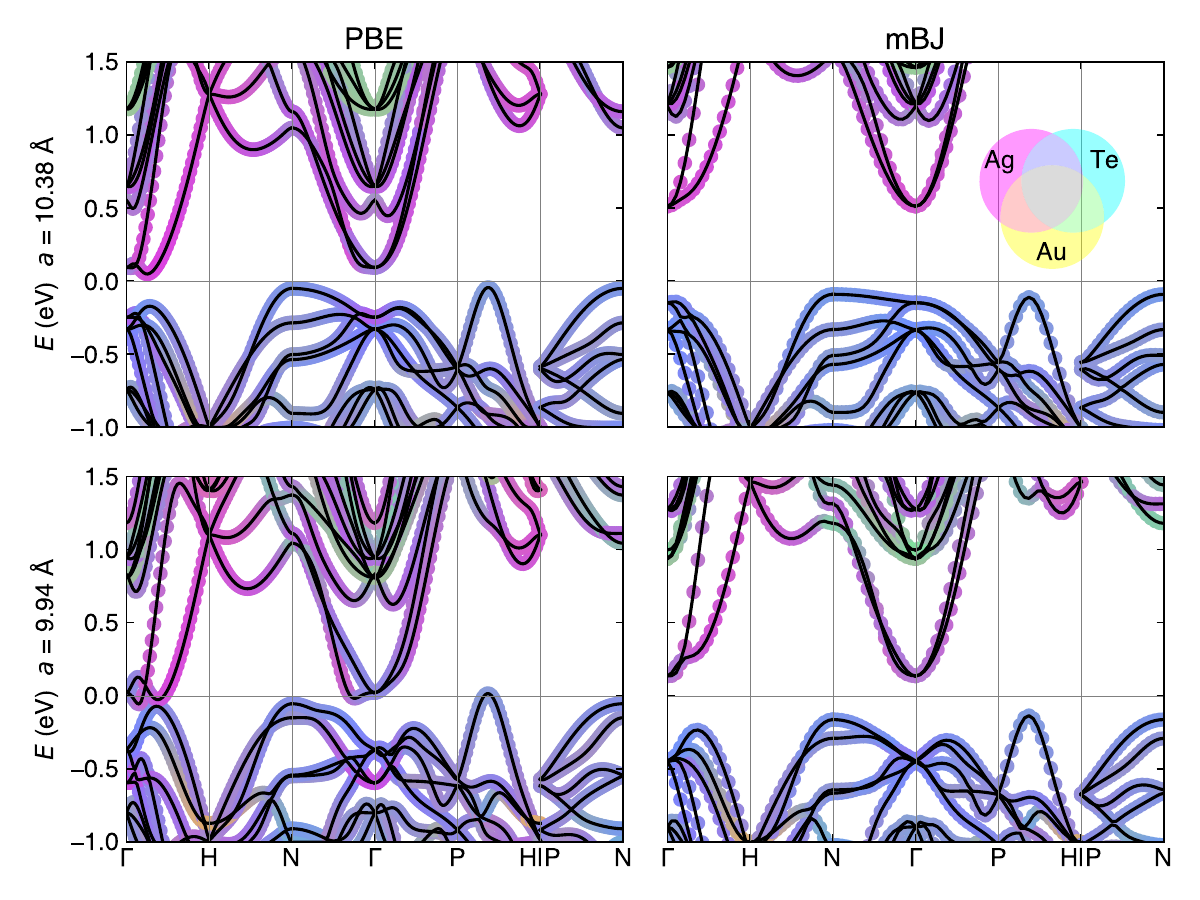} \\
\caption{\label{fig:mbj_wo_soc}
Band structures of Ag$_3$AuTe$_2$ at ambient lattice parameters and -4\% strain (corresponding to 13~GPa) using PBE and mBJ functionals. All band structures were calculated without spin-orbit coupling (SOC). Band structures with SOC are presented in the main manuscript.
}
\end{figure}

\begin{figure}
    \centering
    \includegraphics[width=1\columnwidth]{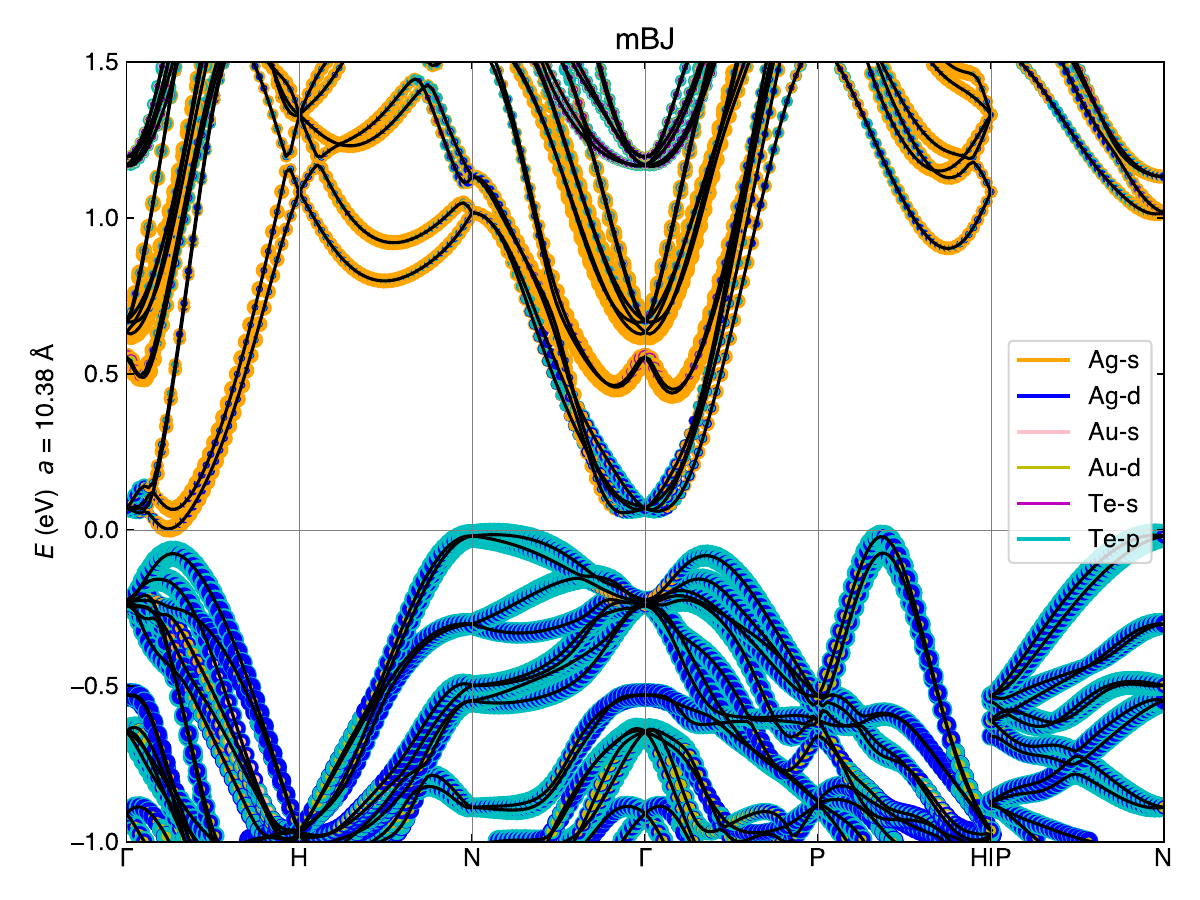}
    \caption{Orbital-projected band structure of Ag$_3$AuTe$_2$ at ambient lattice parameter using PBE functional. The size of the markers is proportional to the orbital contribution. Smaller markers are drawn on top of the larger markers.}
    \label{fig:pbe-soc-ambient}
\end{figure}

\begin{figure}
    \centering
    \includegraphics[width=1\columnwidth]{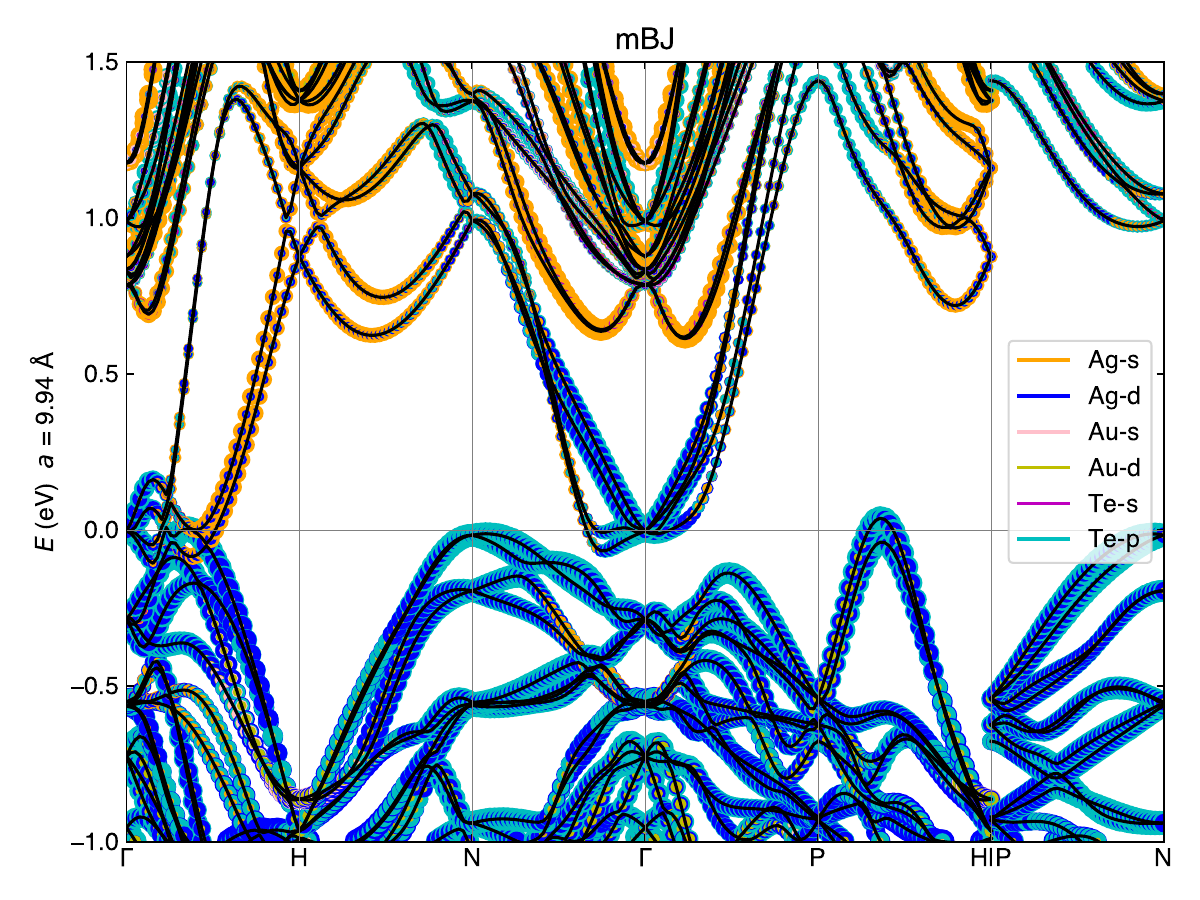}
    \caption{Orbital-projected band structure of Ag$_3$AuTe$_2$ at maximal compressive strain (approximately -4\%) using PBE functional. The size of the markers is proportional to the orbital contribution. Smaller markers are drawn on top of the larger markers.}
    \label{fig:pbe-soc-strain}
\end{figure}

\begin{figure}
    \centering
    \includegraphics[width=1\columnwidth]{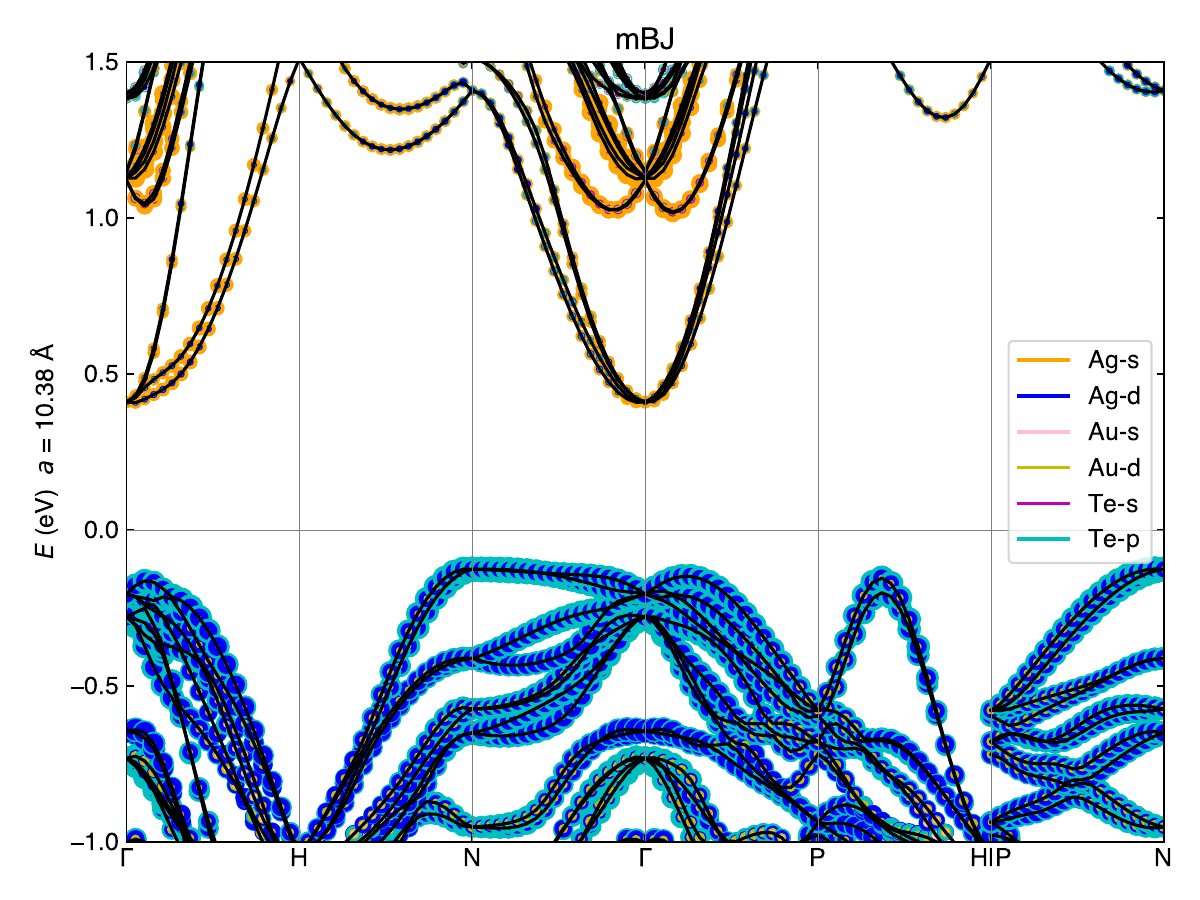}
    \caption{Orbital-projected band structure of Ag$_3$AuTe$_2$ at ambient lattice parameters using mBJ functional. The size of the markers is proportional to the orbital contribution. Smaller markers are drawn on top of the larger markers.}
    \label{fig:mbj-soc-ambient}
\end{figure}

\begin{figure}
    \centering
    \includegraphics[width=1\columnwidth]{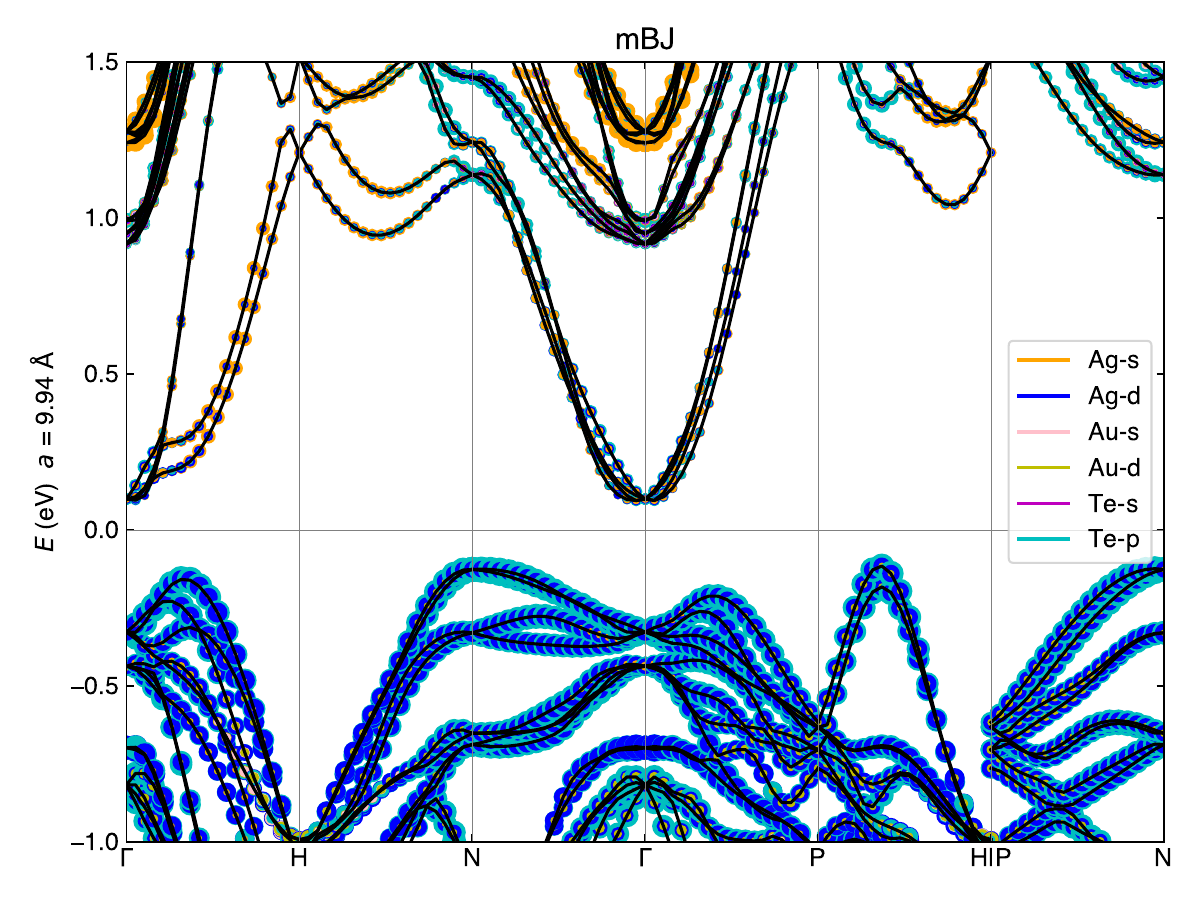}
    \caption{Orbital-projected band structure of Ag$_3$AuTe$_2$ at maximal compressive strain (approximately -4\%) using mBJ functional. The size of the markers is proportional to the orbital contribution. Smaller markers are drawn on top of the larger markers.}
    \label{fig:mbj-soc-strain}
\end{figure}

%\clearpage
%\bibliographystyle{unsrt}
%\bibliography{aat}